\documentclass[journal=nalefd,manuscript=letter]{achemso}
\usepackage{chemformula}
\usepackage[T1]{fontenc}
\usepackage{booktabs}
\usepackage{upgreek}

\author{Lucas E. A. Stehouwer}
\affiliation[QuTech]
{QuTech and Kavli Institute of Nanoscience, Delft University of Technology, Lorentzweg 1, 2628 CJ Delft, The Netherlands}
\author{Merrit P. Losert}
\affiliation{University of Wisconsin-Madison, Madison 53706 WI, USA}
\author{Maia Rigot}
\author{Davide Degli Esposti}
\affiliation[QuTech]
{QuTech and Kavli Institute of Nanoscience, Delft University of Technology, Lorentzweg 1, 2628 CJ Delft, The Netherlands}
\author{Sara Martí-Sánchez}
\affiliation{Catalan Institute of Nanoscience and Nanotechnology (ICN2), CSIC and BIST, Campus UAB, Bellaterra, 08193 Barcelona, Catalonia, Spain}
\author{Maximillian Rimbach-Russ}
\affiliation[QuTech]
{QuTech and Kavli Institute of Nanoscience, Delft University of Technology, Lorentzweg 1, 2628 CJ Delft, The Netherlands}
\author{Jordi Arbiol}
\affiliation{Catalan Institute of Nanoscience and Nanotechnology (ICN2), CSIC and BIST, Campus UAB, Bellaterra, 08193 Barcelona, Catalonia, Spain}
\alsoaffiliation{ICREA, Pg. Lluís Companys 23, 08010 Barcelona, Catalonia, Spain}
\author{Mark Friesen}
\affiliation{University of Wisconsin-Madison, Madison 53706 WI, USA}
\author{Giordano Scappucci}
\email{g.scappucci@tudelft.nl}
\affiliation[QuTech]
{QuTech and Kavli Institute of Nanoscience, Delft University of Technology, Lorentzweg 1, 2628 CJ Delft, The Netherlands}

\date{\today}

\title[An \textsf{achemso} demo]
  {Engineering Ge profiles in Si/SiGe heterostructures for increased valley splitting}

\begin{document}

\begin{abstract}
Electron spin qubits in Si/SiGe quantum wells are limited by the small and variable energy separation of the conduction band valleys. While sharp quantum well interfaces are pursued to increase the valley splitting energy deterministically, here we explore an alternative approach to enhance the valley splitting on average. We grow increasingly thinner quantum wells with broad interfaces to controllably increase the overlap of the electron wave function with Ge atoms. In these quantum wells, comprehensive quantum Hall measurements of two-dimensional electron gases reveal a linear correlation between valley splitting and disorder. Benchmarked against quantum wells with sharp interfaces, we demonstrate enhanced valley splitting while maintaining a low-disorder potential environment. Simulations using the experimental Ge concentration profiles predict an average valley splitting in quantum dots that matches the enhancement observed in two-dimensional systems. Our results motivate the experimental realization of quantum dot spin qubits in these heterostructures.
\end{abstract}

\section{Introduction}

Spin qubit devices in gate-defined Si/SiGe quantum dots have advanced in performance, qubit count, and connectivity. Reproducible single- and two-qubit gate fidelities exceeding 99\% have been achieved~\cite{xue2022quantum, noiri2022fast, mills2022two}.
Moreover, linear array devices have scaled the number of qubits from six~\cite{philips_universal_2022} to twelve~\cite{george_12-spin-qubit_2025}, and a two-by-two qubit array has been demonstrated\cite{unseld2024baseband}.
Coherent, high fidelity spin shuttling~\cite{noiri2022shuttling, seidler2022conveyor, desmet2024high} and cavity-mediated iSWAP oscillations between distant spins~\cite{dijkema_cavity-mediated_2025} are promising achievements for connectivity beyond nearest neighbour.
In addition, the fabrication of Si/SiGe spin qubits in a 300 mm semiconductor manufacturing facility~\cite{neyens_probing_2024} and the integration of multi-level interconnects with 2D spin qubit arrays~\cite{ha_two-dimensional_2025} underscore the potential for scalable architectures.
Despite this compelling progress, critical material challenges remain in the pursuit of a large-scale quantum computer. 

In Si/SiGe heterostructures, a long-standing limitation has been the small and variable energy splitting between the two low-lying conduction band valleys~\cite{friesen_magnetic_2006,zwanenburg_silicon_2013}.
In quantum dots, the reported valley energy splittings vary between a few tens to hundreds of $\upmu$eV \cite{maune2012coherent, zajac2015gate, shi2011tunable, scarlino2017dressed, paquelet2023reducing, mcjunkin2022wiggle, hollmann2020large, borselli2011measurement}.
This poses a challenge for spin qubits, because the increased leakage from the computational two-level Hilbert space affects high-fidelity initialization, control, readout, and shuttling~\cite{vandersypen_interfacing_2017,tagliaferri_impact_2018,huang_spin_2014,seidler_conveyor-mode_2022,langrock_blueprint_2023,zwerver_shuttling_2023,losert2024strategies}. 

Recent work combining experiments and theory\cite{paquelet2022atomic,losert2023practical} has established that the atomistic random alloy concentration fluctuations at the Si/SiGe interface are accountable for the measured valley splitting spread in real quantum dots.
Furthermore, the valley splitting is expected to be enhanced when the electronic wavefunction overlaps with more Ge atoms. While proposed strategies like intentionally adding Ge to the Si quantum well promise increased valley splitting~\cite{paquelet2022atomic,losert2023practical}, they may also worsen disorder, affecting electron mobility~\cite{mcjunkin2022wiggle}.
However, careful tuning of the germanium concentration profile--through adjustments in the Si quantum well thickness, interface width, and barrier composition--can strike a delicate balance between achieving high valley splitting and maintaining low disorder~\cite{degliesposti2024low}.

Here, we engineer the Ge concentration profiles of $^{28}$Si/$^{28}$SiGe heterostructures to enhance the overlap of the electron wave function with Ge atoms in a tunable way, by growing increasingly thin quantum wells with intentionally diffused interfaces.
We characterize the Ge concentration profiles by atomic-resolution scanning transmission electron microscopy (STEM), while we measure the mobility and valley splitting energy of the two-dimensional electron gas ($E_\mathrm{v}$) by comprehensive density-dependent magnetotransport.
Benchmarking against control heterostructures with sharp interfaces~\cite{degliesposti2024low}, we can controllably increase valley splitting by up to a factor of two.
Although we unambiguously observe that higher valley splitting correlates with increased electrical disorder, a beneficial trade-off between enhanced valley splitting and low disorder is achievable. 
Furthermore, simulations of quantum dot valley splitting energy  ($E_\mathrm{v}^\mathrm{QD}$) based on the experimental Ge concentration profiles, reveal a linear relationship with $E_\mathrm{v}$.
This finding provides a first insight into the long-sought connection between valley splitting in the quantum Hall regime and in quantum dots. 

\section{Results and discussion}

\begin{figure}
    \centering
	\includegraphics[width=88mm]{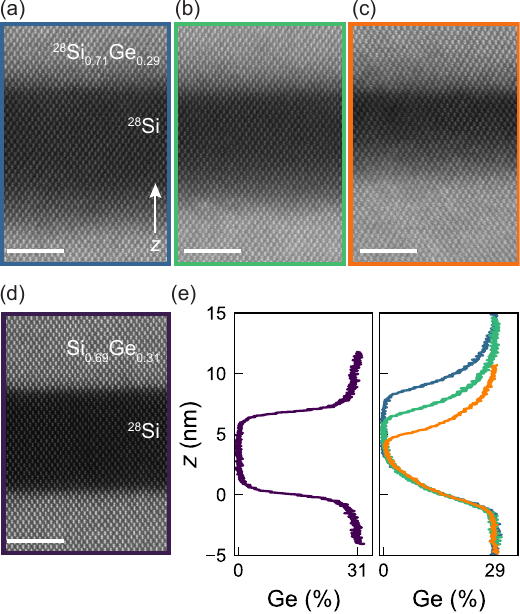}
	\caption 
    {\textbf{$\mathbf{^{28}}$Si/SiGe heterostructures with engineered Ge concentration profiles.} (a)-(c) HAADF-STEM images of $^{28}$Si/SiGe heterostructures B1 (blue), B2 (green), and B3 (orange), with intentionally diffused quantum well interfaces. The quantum well and surrounding barriers feature isotopically purified $^{28}$Si. We vary the quantum well width between B1--B3 from 9.5 to 5.9 nm (see also table\,\ref{tab:one}). (d) HAADF-STEM image of control heterostructure A (purple) with sharp interfaces. Only the quantum well is isotopically purified. Scale bar is 3 nm in (a)--(d). (e) Ge concentration profiles for heterostructure A (left panel) and B1--B3 (right panel) extracted by combining SIMS data with HAADF-STEM intensity profiles (see Fig.\,S1 in the Supporting Information)}
\label{fig:one}
\end{figure}

Figures \ref{fig:one}(a)-(c) show atomic-resolution high angle annular dark field (HAADF) STEM images of three $^{28}$Si/$^{28}$SiGe heterostructures (B1--B3) having progressively thinner quantum wells with similarly broad interfaces. 
As a control, Fig.\,\ref{fig:one}(d) shows $^{28}$Si/SiGe heterostructure (A), with sharp interfaces as studied in \cite{degliesposti2024low}. 
Broad interfaces in heterostructures B1--B3 result from uninterrupted epitaxy at a temperature of 750~\textdegree C using only hydride precursors ($^{28}$SiH$_4$, GeH$_4$). 
In contrast, sharp interfaces in heterostructure A are achieved by growing the SiGe barriers at a lower temperature of 625~\textdegree C, enabled by using a different Si precursor (SiH$_2$Cl$_2$)~\cite{paquelet2022atomic,degliesposti2024low}.
In all heterostructures, the quantum well is deposited on a SiGe strain-relaxed buffer and is separated from the dielectric interface by a 30 nm SiGe barrier (see Section~1 in the Supporting Information). 
Due to the different gas precursors, heterostructures B1--B3 feature an isotopically-enriched barrier with a slightly lower Ge concentration ($^{28}\mathrm{Si}_{0.71}\mathrm{Ge}_{0.29}$) compared to heterostructure A ($\mathrm{Si}_{0.69}\mathrm{Ge}_{0.31}$). The small difference in chemical composition, and therefore band offset, is confirmed by electrical measurements of the quantum well saturation charge density~\cite{lodari_light_2019,degli_esposti_wafer-scale_2022}, which is smaller in B1--B3 compared to A (see Fig.\,S1 in the Supporting Information). 
In Fig.\,\ref{fig:one}(e) we show the Ge concentration profiles from A (left panel) and B1--B3 (right panel) extracted from the HAADF-STEM images (see Fig.\,S1 in the Supporting Information). The right panel highlights both the reproducibility of the growth process from the overlapping bottom interfaces ($z=0$ nm) as well as the control over the quantum well width. From the concentration profiles, we extract the quantum well width $w_\mathrm{QW}$ and the width $w_\mathrm{if}$ of the top and bottom interfaces. Table\,\ref{tab:one} gives a quantitative overview of the extracted parameters. We controllably reduce the quantum well width between the heterostructures B1--B3 by adjusting the quantum well growth time. Notably, the interfaces of heterostructures B1--B3 are approximately 2.4 times wider than those of heterostructure A. 

\begin{table}
\centering
\caption{\textbf{Overview of quantum well metrics.} Quantum well width $w_\mathrm{QW}$, top interface width $w_\mathrm{if}^\mathrm{top}$, and bottom interface width $w_\mathrm{if}^\mathrm{bottom}$ for heterostructures A, B1, B2, and B3 are given. The quantum well width is defined as the distance between top and bottom interface where the Ge concentration reaches 50\% of its maximum value. The width of top and bottom interfaces is defined as the distance over which the Ge concentration rises from 10\% to 90\% of its maximum value. Uncertainty of the extracted values is assumed to be in the last reported digit.}
\setlength{\tabcolsep}{10pt}
\renewcommand{\arraystretch}{1.0}
\begin{tabular}{@{}ccccc@{}}
\midrule
 & \multicolumn{1}{c}{A} & \multicolumn{1}{c}{B1} & \multicolumn{1}{c}{B2} & \multicolumn{1}{c}{B3} \\ \midrule
\multicolumn{1}{c}{$w_\mathrm{QW}$ (nm)} & \multicolumn{1}{c}{6.9} & \multicolumn{1}{c}{9.5} & \multicolumn{1}{c}{7.8} & \multicolumn{1}{c}{5.9} \\ 
\multicolumn{1}{c}{$w_\mathrm{if}^\mathrm{top}$ (nm)} & \multicolumn{1}{c}{1.5} & \multicolumn{1}{c}{3.7} & \multicolumn{1}{c}{3.7} & \multicolumn{1}{c}{3.6} \\ 
\multicolumn{1}{c}{$w_\mathrm{if}^\mathrm{bottom}$ (nm)}& \multicolumn{1}{c}{1.6} & \multicolumn{1}{c}{3.3} & \multicolumn{1}{c}{3.5} & \multicolumn{1}{c}{3.5} \\
\midrule
\end{tabular}%
\label{tab:one}
\end{table}

\begin{figure}
    \centering
	\includegraphics[width=88mm]{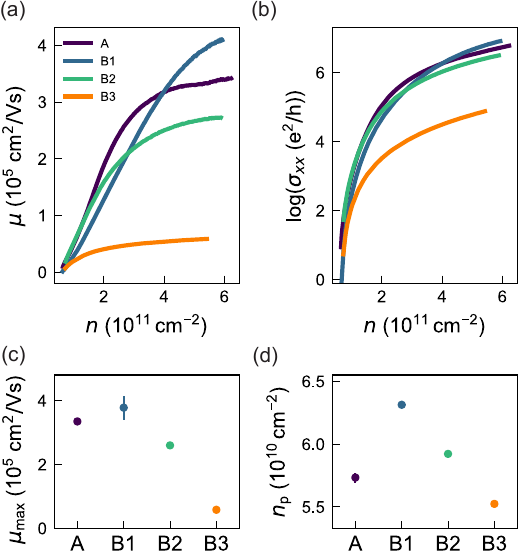}
	\caption{\textbf{Classical transport measurements of disorder.} (a) Mobility ($\mu$)-density ($n$) curves for heterostructures A, B1, B2, and B3. (b) Conductivity ($\sigma_\mathrm{xx}$)-density ($n$) curves of the four heterostructures, from which we fit the percolation density $n_\mathrm{p}$ (see Section~4 in the Supporting Information). (c) Average maximum mobility $\mu_\mathrm{max}$ four the four heterostructures (A1, B1--B3) from measurements of multiple devices. Error bars represent one standard deviation around the average. (d) Extracted percolation densities $n_\mathrm{p}$ for the four heterostructures.}
\label{fig:two}
\end{figure}

We evaluate the disorder properties of the two-dimensional electron gas (2DEG) in each heterostructure by fabricating Hall-bar-shaped heterostructure field effect transistors (H-FETs) and performing magnetotransport measurements at 70~mK in a dilution refrigerator equipped with a cryo-multiplexer~\cite{paquelet2020multiplexed} (see Section~4 in the Supporting Information). Figures \ref{fig:two}(a),(b) show the mobility-density and conductivity-density curves of a representative H-FET for each heterostructure (see Fig.\,S2 in the Supporting Information for other H-FETs). Heterostructures A, B1, and B2 show similar mobility-density curves, while heterostructure B3 shows a severe suppression of the mobility across the entire density range. In Fig.\,\ref{fig:two}(c) we show the average extracted mobility for each heterostructure. Maximum mobility decreases from $3.8(4)\times10^5$ $\mathrm{cm}^2/\mathrm{Vs}$ in B1 to $0.58\times10^5$ 
$\mathrm{cm}^2/\mathrm{Vs}$ in B3 as the quantum well becomes increasingly thinner. Compared to the control heterostructure A, B1 shows a higher average maximum mobility which we attribute to the increased growth temperature resulting in decreased background contamination. However, B1 also shows a larger spread across multiple H-FETs which is indicative of the onset of strain relaxation within the quantum well, creating additional scattering centres from dislocations~\cite{paquelet2023reducing, degliesposti2024low}. The severe reduction of maximum mobility in B3 is compatible with the presence of Ge throughout the thin quantum well. \cite{monroe_comparison_1993,venkataraman_alloy_1993,huang_understanding_2024}. 

In contrast to the observed trend in maximum mobility, we do not observe a strong dependence across different heterostructures of the percolation density $n_\mathrm{p}$ (Fig.\,\ref{fig:two}(d)), obtained by fitting the conductivity curves in Fig.\,\ref{fig:two}(b) (see Section~4 in the Supporting Information). We find similar low values of around $6.0\times10^{10}$ $\mathrm{cm}^{-2}$ for all heterostructures within the constraints of the fitting procedure, which is consistent with previous arguments that alloy disorder only weakly affects the scattering rate at low density \cite{monroe_comparison_1993,venkataraman_alloy_1993,huang_understanding_2024}. This observation suggests that the increased disorder from the diffusion of the interfaces does not severely affect the disorder properties of the 2DEG in the low-density regime, which is relevant for quantum dots. 

\begin{figure}
    \centering
	\includegraphics[width=88mm]{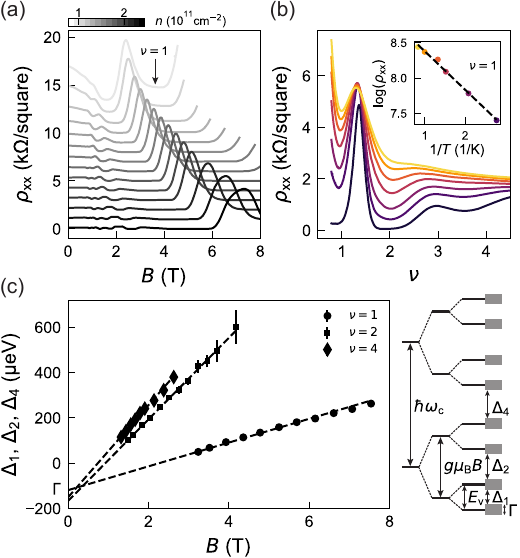}
	\caption{\textbf{Quantum transport measurements of valley splitting.} (a) Longitudinal resistivity $\rho_\mathrm{xx}$ of heterostructure B2 as a function of magnetic field $B$ over a range of fixed densities $n$ between $0.77\times10^{11}\,\,\mathrm{cm}^{-2}$ (light grey) and $2.46\times10^{11}\,\,\mathrm{cm}^{-2}$ (black) (offset for clarity). (b) Magnetotransport measurements at fixed density $n$ (here $0.91\times10^{11}$ cm$^{-2}$) and at different temperatures $T$, plotted against integer filling factor $\nu=nh/eB$. Different colours represent different temperatures between 70 (dark purple) and 1000 mK (yellow). The inset shows the thermally activated dependence of the oscillation minima ($\rho_\mathrm{xx}\propto \exp{(-\Delta/2k_\mathrm{B}T)}$) for integer filling factor $\nu=1$ from which  we extract the mobility gap $\Delta_\mathrm{1}$ of the first valley. These measurements are repeated for each density and the analysis for $\nu =1,2,4$. (c) Mobility gaps of the first valley gap $\Delta_\mathrm{1}$ ($\nu=1$), the first Zeeman gap $\Delta_\mathrm{2}$ ($\nu=2$), and the first Landau gap $\Delta_\mathrm{4}$ ($\nu=4$) as function of magnetic field $B$. The linear fits (dotted lines) are used for extracting the Landau level broadening induced disorder $\Gamma$. The side schematic shows the energy level ladder in the quantum Hall regime, including the disorder broadening $\Gamma$. Landau levels are split by energy $\hbar \omega_\mathrm{c}$, the Zeeman levels by $g\mu_\mathrm{B}B$, and valley levels by the valley splitting energy $E_\mathrm{v}$.}
\label{fig:three}
\end{figure}

After assessing disorder in the heterostructures, we probe valley splitting in the same H-FETs by performing activation energy measurements in the quantum Hall regime, following Ref.\,\cite{paquelet2020effect}\,. We focus on the first valley-split energy gap ($\Delta_1$) at filling factor $\nu=1$, since this gap is resolved across all heterostructures over a similar range of density $n$ and magnetic field $B$, enabling meaningful comparisons. Additionally, we measure the first Zeeman-split gap ($\Delta_\mathrm{2}$), and first Landau gap ($\Delta_\mathrm{4}$), corresponding to $\nu=2$, and $\nu=4$, respectively. Figure\,\ref{fig:three} illustrates the measurement protocol with data from heterostructure B2, while measurements from all other heterostructures are shown in Figs.\,S3--S5 in the Supporting Information. First, we measure the longitudinal resistivity $\rho_\mathrm{xx}$ at base temperature as a function of $B$, over a range of fixed densities $n$ (Fig.~\,\ref{fig:three}(a)). We observe clear Shubnikov de Haas oscillations, with minima at $\nu=1$ reaching zero, indicating a well resolved $\Delta_1$. For each $n$, we repeat the measurement for different temperatures ($T = 70-1000~\mathrm{mK}$) and plot in Fig.\,\ref{fig:three}(b) $\rho_\mathrm{xx}$ as a function of filling factor $\nu$, given by the quantum Hall relation $\nu=nh/eB$, where $h$ is the Planck's constant and $e$ the electron charge. As the inset shows for $\nu=1$, we observe a thermally activated dependence of the oscillation minima ($\rho_\mathrm{xx}\propto \exp{-\Delta/2k_\mathrm{B}T}$). For each density, we extract the valley-split, Zeeman split and Landau mobility gaps ($\Delta_\mathrm{1}$, $\Delta_\mathrm{2}$, $\Delta_\mathrm{4}$ respectively), plotted in Fig.\,\ref{fig:three}(c) as a function of magnetic field $B$. As in Ref.~\cite{paquelet2020effect}\,, we observe striking linear relationships converging to a similar intercept, from which we estimate with confidence the Landau level broadening-induced disorder $\Gamma$ (Fig.\,\ref{fig:three}(c), side panel) and the valley splitting $E_\mathrm{v}=\Delta_\mathrm{1}+\Gamma$.

\begin{figure}
    \centering
	\includegraphics[width=88mm]{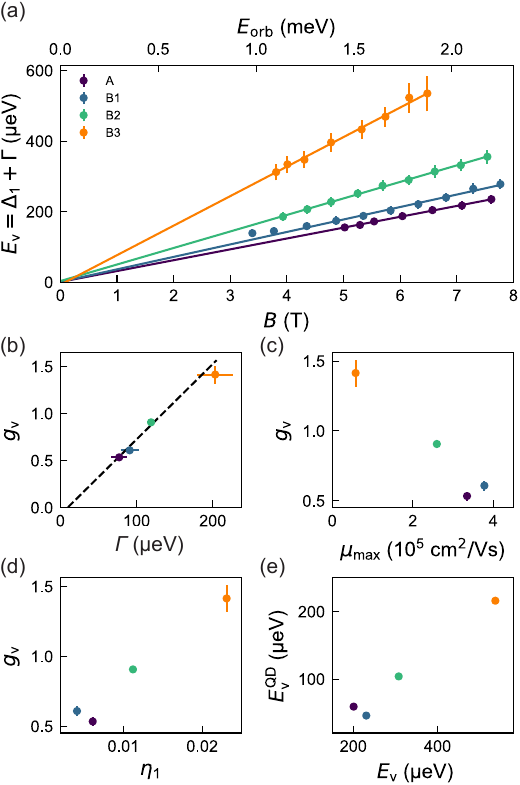}
	\caption{\textbf{Valley splitting correlations} (a) Valley splitting energy $E_\mathrm{v}=\Delta_1+\Gamma$ as a function of magnetic field $B$ for heterostructures A, B1--B3. The corresponding orbital energy $E_{\mathrm{orb}}$ is shown in the top $x$-axis. (b) Valley $g$-factor $g_\mathrm{v}$, from the slopes of $E_\mathrm{v}$ versus magnetic field $B$ in (a), as a function of corresponding Landau level broadening-induced disorder $\Gamma$. Colour coding as in (a). (c) $g_v$ as a function of maximum mobility  $\mu_\mathrm{max}$. Colour coding as in (a). (d) $g_v$ as a function of $\eta_1$, the  overlap of the electron wave function with Ge atoms simulated using the Ge concentration profiles from Fig.\,\ref{fig:one}(e). (d) Simulated valley splitting energy in quantum dots $E_\mathrm{v}^\mathrm{QD}$ against the two-dimensional electron gas valley splitting $E_\mathrm{v}$ from Fig.\,\ref{fig:four}(a) evaluated at a magnetic field $B$ of 6.5 T. This magnetic field corresponds to an orbital energy of 1.88 meV (see main text), which is the same orbital energy used for the simulation of $E_\mathrm{v}^\mathrm{QD}$.}
\label{fig:four}
\end{figure}

Following this systematic classical and quantum transport characterization, we may now investigate the key link between valley splitting and disorder, underpinned by the engineered Ge concentration profiles in the different heterostructures. 
In all heterostructures, $E_\mathrm{v}$ increases linearly with $B$ across the investigated range (Fig.\,\ref{fig:four}(a)).
Additionally, the top $x$-axis shows the correspondingly increasing orbital energy, $E_\mathrm{orb}= e\hbar B/2m^*$, where we use an in-plane effective mass of $0.2m_\mathrm{e}$ for electrons in Si.
Note that the $E_\mathrm{v} \propto B$ relation was previously observed~\cite{paquelet2020effect,goswami_controllable_2007} and attributed~\cite{paquelet2020effect} to the stronger electrostatic confinement achieved for a higher density in the quantum Hall edge channel, driven by $B$ via the quantum Hall relation $n=eB/h$ for $\nu=1$.

Across the explored magnetic field range, we observe a clear trend in Fig.\,\ref{fig:four}(a): all heterostructures with broad interfaces (B1--B3) show larger valley splitting compared to the control heterostructure A. Moreover, within heterostructures B1--B3, thinner quantum wells achieve larger valley splitting, validating our heterostructure design. 
To quantify these observations, we extract the valley $g$-factor, $g_\mathrm{v}=(1/\mu_\mathrm{B})(\mathrm{d}E_\mathrm{v}/\mathrm{d}B)$, which represents the rate of change of valley splitting with magnetic field, normalized to the Bohr magneton $\mu_\mathrm{B}$.
Figure\,\ref{fig:four}(b) shows $g_\mathrm{v}$ against the disorder-induced Landau level broadening $\Gamma$, revealing a striking experimental correlation, further corroborated in Figure\,\ref{fig:four}(c) by the dependence of $g_\mathrm{v}$ against the maximum mobility. The valley splitting in heterostructures B1--B3 may increase to more than twice the value observed in the control heterostructure A.
These clear trends confirm the intuition that increasing valley splitting, which requires breaking translation symmetry, comes at the expense of a more disordered potential landscape~\cite{losert2023practical,degliesposti2024low}, qualified in our experiments either by classical or quantum transport measurements. 

Next, we investigate the atomistic origin of the increased valley splitting $E_\mathrm{v}$ and provide a prognosis for potential gains for valley splitting in quantum dots. To this end, we calculate for each heterostructure the parameter $\eta_1$ (see Section~5 in the Supporting Information), which quantifies the overlap of the electron wave function with Ge atoms, and simulate the quantum dot valley splitting distributions and their mean value $E_\mathrm{v}^\mathrm{QD}$. We use the extracted Ge concentration profiles of Fig.\,\ref{fig:one}(e) as an experimental input for the simulation methods in Ref.\,\cite{losert2023practical}\,.
As Fig.\,\ref{fig:four}(d) shows, we find an unambiguous correlation between $g_\mathrm{v}$ and $\eta_1$, suggesting that the larger $E_\mathrm{v}$ measured in the 2DEG correlates with increased overlap of the electron wave function with Ge atoms, promoted in our experiments by thinner quantum wells with broad interfaces. This finding ($E_\mathrm{v} \propto \eta_1$) mirrors the theoretical predictions for average valley splitting in alloy disorder-dominated quantum dots ($E_\mathrm{v}^\mathrm{QD} \propto \eta_1$ (see Section~5 in the Supporting Information and Refs.\cite{paquelet2022atomic,losert2023practical}). As a consequence, the plot in Fig.\,\ref{fig:four}(e) of simulated $E_\mathrm{v}^\mathrm{QD}$ against experimentally measured $E_\mathrm{v}$ shows also a linear relationship. Here,
we choose to simulate $E_\mathrm{v}^\mathrm{QD}$ at an orbital energy of 1.88 meV, which is on par with measured values in quantum dots and corresponds from Fig.\,\ref{fig:four}(a) to an experimentally accessible magnetic field of 6.5 T for evaluation of $E_\mathrm{v}$. 
Considering the same orbital energy ensures a meaningful comparison between the experimentally-informed simulation of $E_\mathrm{v}^\mathrm{QD}$ and the measured $E_\mathrm{v}$. 
While offering a first insight into the relation between the two metrics, based on these results we predict that heterostructures B1--B3 could support on average increased valley splitting in quantum dots, which is proxied by valley splitting measured in the quantum Hall regime.

\section{Conclusions}
In summary, we have engineered $^{28}\mathrm{Si}/\mathrm{^{28}SiGe}$ heterostructures to enhance the overlap of the electron wave function with germanium atoms, by growing increasingly thin quantum wells with intentionally diffused interfaces. Our comprehensive study unveils unambiguously a correlation between disorder in the two-dimensional electron gas, measured with classical and quantum transport, and valley splitting, measured in the quantum Hall regime. Valley splitting is increased but so is disorder. Based on simulations that take into account the experimental Ge concentration profile, we identify the overlap of the electron wavefunction with Ge atoms as the likely cause of this connection, which also propagates to calculated average values of valley splitting distributions in quantum dots.

Compared to control samples with sharp interfaces in Ref.~\cite{degliesposti2024low}\,, we show that a quantum well with much broader interfaces ($\simeq$3.6~nm) and similar width ($\simeq 7.8$~nm) offers an excellent trade-off, featuring a $1.8\times$ valley splitting increase, whilst still having respectable mobility ($>2\times10^5$~cm$^2$/Vs) and low percolation density ($<6\times10^{10}$~cm$^{-2}$). In contrast, thinner or thicker quantum wells either significantly degrade mobility or yield only marginal improvements in valley splitting.    
Future statistical studies of valley splitting in quantum dots fabricated on these new generations of heterostructures are required to confirm the valley splitting increase and  assess the impact of random alloy disorder in the formation of spurious quantum dots.

\section{Acknowledgments}
This work was supported by the Netherlands Organisation for Scientific Research (NWO/OCW), via the Frontiers of Nanoscience program Open Competition Domain Science - M program.
We acknowledge support by the European Union through the IGNITE project with grant agreement No. 101069515 and the QLSI project with grant agreement No. 951852.
This research was sponsored in part by the Army Research Office (ARO) under Awards No. W911NF-23-1-0110 and W911NF-22-1-0090 . The views, conclusions, and recommendations contained in this document are those of the authors and are not necessarily endorsed nor should they be interpreted as representing the official policies, either expressed or implied, of the Army Research Office (ARO) or the U.S. Government. The U.S. Government is authorized to reproduce and distribute reprints for Government purposes notwithstanding any copyright notation herein.
This research was sponsored in part by The Netherlands Ministry of Defence under Awards No. QuBits R23/009. The views, conclusions, and recommendations contained in this document are those of the authors and are not necessarily endorsed nor should they be interpreted as representing the official policies, either expressed or implied, of The Netherlands Ministry of Defence. The Netherlands Ministry of Defence is authorized to reproduce and distribute reprints for Government purposes notwithstanding any copyright notation herein.
ICN2 acknowledges funding from Generalitat de Catalunya 2021SGR00457. We acknowledge support from CSIC Interdisciplinary Thematic Platform (PTI+) on Quantum Technologies (PTI-QTEP+). This research work has been funded by the European Commission – NextGenerationEU (Regulation EU 2020/2094), through CSIC's Quantum Technologies Platform (QTEP). ICN2 is supported by the Severo Ochoa program from Spanish MCIN / AEI (Grant No.: CEX2021-001214-S) and is funded by the CERCA Programme / Generalitat de Catalunya. Authors acknowledge the use of instrumentation as well as the technical advice provided by the Joint Electron Microscopy Center at ALBA (JEMCA). ICN2 acknowledges funding from Grant IU16-014206 (METCAM-FIB) funded by the European Union through the European Regional Development Fund (ERDF), with the support of the Ministry of Research and Universities, Generalitat de Catalunya.

\section*{Data availability}
The data sets supporting the findings of this study are openly available in 4TU Research Data at \url{https://doi.org/10.4121/ebcf5563-628e-479c-9e0d-d5094ebb9c27}.

\bibliography{achemso-demo}

\providecommand{\latin}[1]{#1}
\makeatletter
\providecommand{\doi}
  {\begingroup\let\do\@makeother\dospecials
  \catcode`\{=1 \catcode`\}=2 \doi@aux}
\providecommand{\doi@aux}[1]{\endgroup\texttt{#1}}
\makeatother
\providecommand*\mcitethebibliography{\thebibliography}
\csname @ifundefined\endcsname{endmcitethebibliography}  {\let\endmcitethebibliography\endthebibliography}{}
\begin{mcitethebibliography}{41}
\providecommand*\natexlab[1]{#1}
\providecommand*\mciteSetBstSublistMode[1]{}
\providecommand*\mciteSetBstMaxWidthForm[2]{}
\providecommand*\mciteBstWouldAddEndPuncttrue
  {\def\EndOfBibitem{\unskip.}}
\providecommand*\mciteBstWouldAddEndPunctfalse
  {\let\EndOfBibitem\relax}
\providecommand*\mciteSetBstMidEndSepPunct[3]{}
\providecommand*\mciteSetBstSublistLabelBeginEnd[3]{}
\providecommand*\EndOfBibitem{}
\mciteSetBstSublistMode{f}
\mciteSetBstMaxWidthForm{subitem}{(\alph{mcitesubitemcount})}
\mciteSetBstSublistLabelBeginEnd
  {\mcitemaxwidthsubitemform\space}
  {\relax}
  {\relax}

\bibitem[Xue \latin{et~al.}(2022)Xue, Russ, Samkharadze, Undseth, Sammak, Scappucci, and Vandersypen]{xue2022quantum}
Xue,~X.; Russ,~M.; Samkharadze,~N.; Undseth,~B.; Sammak,~A.; Scappucci,~G.; Vandersypen,~L. M.~K. Quantum logic with spin qubits crossing the surface code threshold. \emph{Nature} \textbf{2022}, \emph{601}, 343--347\relax
\mciteBstWouldAddEndPuncttrue
\mciteSetBstMidEndSepPunct{\mcitedefaultmidpunct}
{\mcitedefaultendpunct}{\mcitedefaultseppunct}\relax
\EndOfBibitem
\bibitem[Noiri \latin{et~al.}(2022)Noiri, Takeda, Nakajima, Kobayashi, Sammak, Scappucci, and Tarucha]{noiri2022fast}
Noiri,~A.; Takeda,~K.; Nakajima,~T.; Kobayashi,~T.; Sammak,~A.; Scappucci,~G.; Tarucha,~S. Fast universal quantum gate above the fault-tolerance threshold in silicon. \emph{Nature} \textbf{2022}, \emph{601}, 338--342\relax
\mciteBstWouldAddEndPuncttrue
\mciteSetBstMidEndSepPunct{\mcitedefaultmidpunct}
{\mcitedefaultendpunct}{\mcitedefaultseppunct}\relax
\EndOfBibitem
\bibitem[Mills \latin{et~al.}(2022)Mills, Guinn, Gullans, Sigillito, Feldman, Nielsen, and Petta]{mills2022two}
Mills,~A.~R.; Guinn,~C.~R.; Gullans,~M.~J.; Sigillito,~A.~J.; Feldman,~M.~M.; Nielsen,~E.; Petta,~J.~R. Two-qubit silicon quantum processor with operation fidelity exceeding 99
\mciteBstWouldAddEndPuncttrue
\mciteSetBstMidEndSepPunct{\mcitedefaultmidpunct}
{\mcitedefaultendpunct}{\mcitedefaultseppunct}\relax
\EndOfBibitem
\bibitem[Philips \latin{et~al.}(2022)Philips, Madzik, Amitonov, de~Snoo, Russ, Kalhor, Volk, Lawrie, Brousse, Tryputen, Wuetz, Sammak, Veldhorst, Scappucci, and Vandersypen]{philips_universal_2022}
Philips,~S. G.~J.; Madzik,~M.~T.; Amitonov,~S.~V.; de~Snoo,~S.~L.; Russ,~M.; Kalhor,~N.; Volk,~C.; Lawrie,~W. I.~L.; Brousse,~D.; Tryputen,~L.; Wuetz,~B.~P.; Sammak,~A.; Veldhorst,~M.; Scappucci,~G.; Vandersypen,~L. M.~K. Universal control of a six-qubit quantum processor in silicon. \emph{Nature} \textbf{2022}, \emph{609}, 919--924\relax
\mciteBstWouldAddEndPuncttrue
\mciteSetBstMidEndSepPunct{\mcitedefaultmidpunct}
{\mcitedefaultendpunct}{\mcitedefaultseppunct}\relax
\EndOfBibitem
\bibitem[George \latin{et~al.}(2025)George, Mądzik, Henry, Wagner, Islam, Borjans, Connors, Corrigan, Curry, Harper, Keith, Lampert, Luthi, Mohiyaddin, Murcia, Nair, Nahm, Nethwewala, Neyens, Patra, Raharjo, Rogan, Savytskyy, Watson, Ziegler, Zietz, Pellerano, Pillarisetty, Bishop, Bojarski, Roberts, and Clarke]{george_12-spin-qubit_2025}
George,~H.~C. \latin{et~al.}  12-{Spin}-{Qubit} {Arrays} {Fabricated} on a 300 mm {Semiconductor} {Manufacturing} {Line}. \emph{Nano Letters} \textbf{2025}, \emph{25}, 793--799\relax
\mciteBstWouldAddEndPuncttrue
\mciteSetBstMidEndSepPunct{\mcitedefaultmidpunct}
{\mcitedefaultendpunct}{\mcitedefaultseppunct}\relax
\EndOfBibitem
\bibitem[Unseld \latin{et~al.}(2024)Unseld, Undseth, Raymenants, Matsumoto, Karwal, Pietx-Casas, Ivlev, Meyer, Sammak, Veldhorst, Scappucci, and Vandersypen]{unseld2024baseband}
Unseld,~F.~K.; Undseth,~B.; Raymenants,~E.; Matsumoto,~Y.; Karwal,~S.; Pietx-Casas,~O.; Ivlev,~A.~S.; Meyer,~M.; Sammak,~A.; Veldhorst,~M.; Scappucci,~G.; Vandersypen,~L. M.~K. Baseband control of single-electron silicon spin qubits in two dimensions. 2024\relax
\mciteBstWouldAddEndPuncttrue
\mciteSetBstMidEndSepPunct{\mcitedefaultmidpunct}
{\mcitedefaultendpunct}{\mcitedefaultseppunct}\relax
\EndOfBibitem
\bibitem[Noiri \latin{et~al.}(2022)Noiri, Takeda, Nakajima, Kobayashi, Sammak, Scappucci, and Tarucha]{noiri2022shuttling}
Noiri,~A.; Takeda,~K.; Nakajima,~T.; Kobayashi,~T.; Sammak,~A.; Scappucci,~G.; Tarucha,~S. A shuttling-based two-qubit logic gate for linking distant silicon quantum processors. \emph{Nature Communications} \textbf{2022}, \emph{13}\relax
\mciteBstWouldAddEndPuncttrue
\mciteSetBstMidEndSepPunct{\mcitedefaultmidpunct}
{\mcitedefaultendpunct}{\mcitedefaultseppunct}\relax
\EndOfBibitem
\bibitem[Seidler \latin{et~al.}(2022)Seidler, Struck, Xue, Focke, Trellenkamp, Bluhm, and Schreiber]{seidler2022conveyor}
Seidler,~I.; Struck,~T.; Xue,~R.; Focke,~N.; Trellenkamp,~S.; Bluhm,~H.; Schreiber,~L.~R. Conveyor-mode single-electron shuttling in Si/SiGe for a scalable quantum computing architecture. \emph{npj Quantum Information} \textbf{2022}, \emph{8}\relax
\mciteBstWouldAddEndPuncttrue
\mciteSetBstMidEndSepPunct{\mcitedefaultmidpunct}
{\mcitedefaultendpunct}{\mcitedefaultseppunct}\relax
\EndOfBibitem
\bibitem[De~Smet \latin{et~al.}(2024)De~Smet, Matsumoto, Zwerver, Tryputen, de~Snoo, Amitonov, Sammak, Samkharadze, G{\"u}l, Wasserman, Rimbach-Russ, Scappucci, and Vandersypen]{desmet2024high}
De~Smet,~M.; Matsumoto,~Y.; Zwerver,~A.-M.~J.; Tryputen,~L.; de~Snoo,~S.~L.; Amitonov,~S.~V.; Sammak,~A.; Samkharadze,~N.; G{\"u}l,~{\"O}.; Wasserman,~R. N.~M.; Rimbach-Russ,~M.; Scappucci,~G.; Vandersypen,~L. M.~K. High-fidelity single-spin shuttling in silicon. 2024\relax
\mciteBstWouldAddEndPuncttrue
\mciteSetBstMidEndSepPunct{\mcitedefaultmidpunct}
{\mcitedefaultendpunct}{\mcitedefaultseppunct}\relax
\EndOfBibitem
\bibitem[Dijkema \latin{et~al.}(2025)Dijkema, Xue, Harvey-Collard, Rimbach-Russ, de~Snoo, Zheng, Sammak, Scappucci, and Vandersypen]{dijkema_cavity-mediated_2025}
Dijkema,~J.; Xue,~X.; Harvey-Collard,~P.; Rimbach-Russ,~M.; de~Snoo,~S.~L.; Zheng,~G.; Sammak,~A.; Scappucci,~G.; Vandersypen,~L. M.~K. Cavity-mediated {iSWAP} oscillations between distant spins. \emph{Nature Physics} \textbf{2025}, \emph{21}, 168--174\relax
\mciteBstWouldAddEndPuncttrue
\mciteSetBstMidEndSepPunct{\mcitedefaultmidpunct}
{\mcitedefaultendpunct}{\mcitedefaultseppunct}\relax
\EndOfBibitem
\bibitem[Neyens \latin{et~al.}(2024)Neyens, Zietz, Watson, Luthi, Nethwewala, George, Henry, Islam, Wagner, Borjans, Connors, Corrigan, Curry, Keith, Kotlyar, Lampert, Mądzik, Millard, Mohiyaddin, Pellerano, Pillarisetty, Ramsey, Savytskyy, Schaal, Zheng, Ziegler, Bishop, Bojarski, Roberts, and Clarke]{neyens_probing_2024}
Neyens,~S. \latin{et~al.}  Probing single electrons across 300-mm spin qubit wafers. \emph{Nature} \textbf{2024}, \emph{629}, 80--85\relax
\mciteBstWouldAddEndPuncttrue
\mciteSetBstMidEndSepPunct{\mcitedefaultmidpunct}
{\mcitedefaultendpunct}{\mcitedefaultseppunct}\relax
\EndOfBibitem
\bibitem[Ha \latin{et~al.}(2025)Ha, Acuna, Raach, Bloom, Brecht, Chappell, Choi, Christensen, Counts, Daprano, Dodson, Eng, Fialkow, Garcia, Ha, Harris, holman, Khalaf, Matten, Peterson, Plesha, Ruiz, Smith, Thomas, Whiteley, Ladd, Jura, Rakher, and Borselli]{ha_two-dimensional_2025}
Ha,~S.~D. \latin{et~al.}  Two-dimensional {Si} spin qubit arrays with multilevel interconnects. 2025; \url{http://arxiv.org/abs/2502.08861}, arXiv:2502.08861\relax
\mciteBstWouldAddEndPuncttrue
\mciteSetBstMidEndSepPunct{\mcitedefaultmidpunct}
{\mcitedefaultendpunct}{\mcitedefaultseppunct}\relax
\EndOfBibitem
\bibitem[Friesen \latin{et~al.}(2006)Friesen, Eriksson, and Coppersmith]{friesen_magnetic_2006}
Friesen,~M.; Eriksson,~M.~A.; Coppersmith,~S.~N. Magnetic field dependence of valley splitting in realistic Si/SiGe quantum wells. \emph{Applied Physics Letters} \textbf{2006}, \emph{89}, 202106\relax
\mciteBstWouldAddEndPuncttrue
\mciteSetBstMidEndSepPunct{\mcitedefaultmidpunct}
{\mcitedefaultendpunct}{\mcitedefaultseppunct}\relax
\EndOfBibitem
\bibitem[Zwanenburg \latin{et~al.}(2013)Zwanenburg, Dzurak, Morello, Simmons, Hollenberg, Klimeck, Rogge, Coppersmith, and Eriksson]{zwanenburg_silicon_2013}
Zwanenburg,~F.~A.; Dzurak,~A.~S.; Morello,~A.; Simmons,~M.~Y.; Hollenberg,~L. C.~L.; Klimeck,~G.; Rogge,~S.; Coppersmith,~S.~N.; Eriksson,~M.~A. Silicon quantum electronics. \emph{Rev. Mod. Phys.} \textbf{2013}, \emph{85}, 961--1019\relax
\mciteBstWouldAddEndPuncttrue
\mciteSetBstMidEndSepPunct{\mcitedefaultmidpunct}
{\mcitedefaultendpunct}{\mcitedefaultseppunct}\relax
\EndOfBibitem
\bibitem[Maune \latin{et~al.}(2012)Maune, Borselli, Huang, Ladd, Deelman, Holabird, Kiselev, Alvarado-Rodriguez, Ross, Schmitz, Sokolich, Watson, Gyure, and Hunter]{maune2012coherent}
Maune,~B.~M.; Borselli,~M.~G.; Huang,~B.; Ladd,~T.~D.; Deelman,~P.~W.; Holabird,~K.~S.; Kiselev,~A.~A.; Alvarado-Rodriguez,~I.; Ross,~R.~S.; Schmitz,~A.~E.; Sokolich,~M.; Watson,~C.~A.; Gyure,~M.~F.; Hunter,~A.~T. Coherent singlet-triplet oscillations in a silicon-based double quantum dot. \emph{Nature} \textbf{2012}, \emph{481}, 344--347\relax
\mciteBstWouldAddEndPuncttrue
\mciteSetBstMidEndSepPunct{\mcitedefaultmidpunct}
{\mcitedefaultendpunct}{\mcitedefaultseppunct}\relax
\EndOfBibitem
\bibitem[Zajac \latin{et~al.}(2015)Zajac, Hazard, Mi, Wang, and Petta]{zajac2015gate}
Zajac,~D.~M.; Hazard,~T.~M.; Mi,~X.; Wang,~K.; Petta,~J.~R. A reconfigurable gate architecture for Si/SiGe quantum dots. \emph{Applied Physics Letters} \textbf{2015}, \emph{106}\relax
\mciteBstWouldAddEndPuncttrue
\mciteSetBstMidEndSepPunct{\mcitedefaultmidpunct}
{\mcitedefaultendpunct}{\mcitedefaultseppunct}\relax
\EndOfBibitem
\bibitem[Shi \latin{et~al.}(2011)Shi, Simmons, Prance, King~Gamble, Friesen, Savage, Lagally, Coppersmith, and Eriksson]{shi2011tunable}
Shi,~Z.; Simmons,~C.~B.; Prance,~J.~R.; King~Gamble,~J.; Friesen,~M.; Savage,~D.~E.; Lagally,~M.~G.; Coppersmith,~S.~N.; Eriksson,~M.~A. Tunable singlet-triplet splitting in a few-electron Si/SiGe quantum dot. \emph{Applied Physics Letters} \textbf{2011}, \emph{99}\relax
\mciteBstWouldAddEndPuncttrue
\mciteSetBstMidEndSepPunct{\mcitedefaultmidpunct}
{\mcitedefaultendpunct}{\mcitedefaultseppunct}\relax
\EndOfBibitem
\bibitem[Scarlino \latin{et~al.}(2017)Scarlino, Kawakami, Jullien, Ward, Savage, Lagally, Friesen, Coppersmith, Eriksson, and Vandersypen]{scarlino2017dressed}
Scarlino,~P.; Kawakami,~E.; Jullien,~T.; Ward,~D.~R.; Savage,~D.~E.; Lagally,~M.~G.; Friesen,~M.; Coppersmith,~S.~N.; Eriksson,~M.~A.; Vandersypen,~L. M.~K. Dressed photon-orbital states in a quantum dot: Intervalley spin resonance. \emph{Physical Review B} \textbf{2017}, \emph{95}, 165429\relax
\mciteBstWouldAddEndPuncttrue
\mciteSetBstMidEndSepPunct{\mcitedefaultmidpunct}
{\mcitedefaultendpunct}{\mcitedefaultseppunct}\relax
\EndOfBibitem
\bibitem[Paquelet~Wuetz \latin{et~al.}(2023)Paquelet~Wuetz, Degli~Esposti, Zwerver, Amitonov, Botifoll, Arbiol, Vandersypen, Russ, and Scappucci]{paquelet2023reducing}
Paquelet~Wuetz,~B.; Degli~Esposti,~D.; Zwerver,~A.-M.~J.; Amitonov,~S.~V.; Botifoll,~M.; Arbiol,~J.; Vandersypen,~L. M.~K.; Russ,~M.; Scappucci,~G. Reducing charge noise in quantum dots by using thin silicon quantum wells. \emph{Nature Communications} \textbf{2023}, \emph{14}\relax
\mciteBstWouldAddEndPuncttrue
\mciteSetBstMidEndSepPunct{\mcitedefaultmidpunct}
{\mcitedefaultendpunct}{\mcitedefaultseppunct}\relax
\EndOfBibitem
\bibitem[McJunkin \latin{et~al.}(2022)McJunkin, Harpt, Feng, Losert, Rahman, Dodson, Wolfe, Savage, Lagally, Coppersmith, Friesen, Joynt, and Eriksson]{mcjunkin2022wiggle}
McJunkin,~T.; Harpt,~B.; Feng,~Y.; Losert,~M.~P.; Rahman,~R.; Dodson,~J.~P.; Wolfe,~M.~A.; Savage,~D.~E.; Lagally,~M.~G.; Coppersmith,~S.~N.; Friesen,~M.; Joynt,~R.; Eriksson,~M.~A. SiGe quantum wells with oscillating Ge concentrations for quantum dot qubits. \emph{Nature Communications} \textbf{2022}, \emph{13}\relax
\mciteBstWouldAddEndPuncttrue
\mciteSetBstMidEndSepPunct{\mcitedefaultmidpunct}
{\mcitedefaultendpunct}{\mcitedefaultseppunct}\relax
\EndOfBibitem
\bibitem[Hollmann \latin{et~al.}(2020)Hollmann, Struck, Langrock, Schmidbauer, Schauer, Leonhardt, Sawano, Riemann, Abrosimov, Bougeard, and Schreiber]{hollmann2020large}
Hollmann,~A.; Struck,~T.; Langrock,~V.; Schmidbauer,~A.; Schauer,~F.; Leonhardt,~T.; Sawano,~K.; Riemann,~H.; Abrosimov,~N.~V.; Bougeard,~D.; Schreiber,~L.~R. Large, Tunable Valley Splitting and Single-Spin Relaxation Mechanisms in a {Si/$\mathrm{Si}_{x}\mathrm{Ge}_{1-x}$} Quantum Dot. \emph{Physical Review Applied} \textbf{2020}, \emph{13}, 034068\relax
\mciteBstWouldAddEndPuncttrue
\mciteSetBstMidEndSepPunct{\mcitedefaultmidpunct}
{\mcitedefaultendpunct}{\mcitedefaultseppunct}\relax
\EndOfBibitem
\bibitem[Borselli \latin{et~al.}(2011)Borselli, Ross, Kiselev, Croke, Holabird, Deelman, Warren, Alvarado-Rodriguez, Milosavljevic, Ku, Wong, Schmitz, Sokolich, Gyure, and Hunter]{borselli2011measurement}
Borselli,~M.~G.; Ross,~R.~S.; Kiselev,~A.~A.; Croke,~E.~T.; Holabird,~K.~S.; Deelman,~P.~W.; Warren,~L.~D.; Alvarado-Rodriguez,~I.; Milosavljevic,~I.; Ku,~F.~C.; Wong,~W.~S.; Schmitz,~A.~E.; Sokolich,~M.; Gyure,~M.~F.; Hunter,~A.~T. Measurement of valley splitting in high-symmetry Si/SiGe quantum dots. \emph{Applied Physics Letters} \textbf{2011}, \emph{98}\relax
\mciteBstWouldAddEndPuncttrue
\mciteSetBstMidEndSepPunct{\mcitedefaultmidpunct}
{\mcitedefaultendpunct}{\mcitedefaultseppunct}\relax
\EndOfBibitem
\bibitem[Vandersypen \latin{et~al.}(2017)Vandersypen, Bluhm, Clarke, Dzurak, Ishihara, Morello, Reilly, Schreiber, and Veldhorst]{vandersypen_interfacing_2017}
Vandersypen,~L. M.~K.; Bluhm,~H.; Clarke,~J.~S.; Dzurak,~A.~S.; Ishihara,~R.; Morello,~A.; Reilly,~D.~J.; Schreiber,~L.~R.; Veldhorst,~M. Interfacing spin qubits in quantum dots and donors—hot, dense, and coherent. \emph{npj Quantum Information} \textbf{2017}, \emph{3}, 1--10\relax
\mciteBstWouldAddEndPuncttrue
\mciteSetBstMidEndSepPunct{\mcitedefaultmidpunct}
{\mcitedefaultendpunct}{\mcitedefaultseppunct}\relax
\EndOfBibitem
\bibitem[Tagliaferri \latin{et~al.}(2018)Tagliaferri, Bavdaz, Huang, Dzurak, Culcer, and Veldhorst]{tagliaferri_impact_2018}
Tagliaferri,~M. L.~V.; Bavdaz,~P.~L.; Huang,~W.; Dzurak,~A.~S.; Culcer,~D.; Veldhorst,~M. Impact of valley phase and splitting on readout of silicon spin qubits. \emph{Physical Review B} \textbf{2018}, \emph{97}, 245412\relax
\mciteBstWouldAddEndPuncttrue
\mciteSetBstMidEndSepPunct{\mcitedefaultmidpunct}
{\mcitedefaultendpunct}{\mcitedefaultseppunct}\relax
\EndOfBibitem
\bibitem[Huang and Hu(2014)Huang, and Hu]{huang_spin_2014}
Huang,~P.; Hu,~X. Spin relaxation in a {Si} quantum dot due to spin-valley mixing. \emph{Physical Review B} \textbf{2014}, \emph{90}, 235315\relax
\mciteBstWouldAddEndPuncttrue
\mciteSetBstMidEndSepPunct{\mcitedefaultmidpunct}
{\mcitedefaultendpunct}{\mcitedefaultseppunct}\relax
\EndOfBibitem
\bibitem[Seidler \latin{et~al.}(2022)Seidler, Struck, Xue, Focke, Trellenkamp, Bluhm, and Schreiber]{seidler_conveyor-mode_2022}
Seidler,~I.; Struck,~T.; Xue,~R.; Focke,~N.; Trellenkamp,~S.; Bluhm,~H.; Schreiber,~L.~R. Conveyor-mode single-electron shuttling in {Si}/{SiGe} for a scalable quantum computing architecture. \emph{npj Quantum Information} \textbf{2022}, \emph{8}, 100\relax
\mciteBstWouldAddEndPuncttrue
\mciteSetBstMidEndSepPunct{\mcitedefaultmidpunct}
{\mcitedefaultendpunct}{\mcitedefaultseppunct}\relax
\EndOfBibitem
\bibitem[Langrock \latin{et~al.}(2023)Langrock, Krzywda, Focke, Seidler, Schreiber, and Cywinski]{langrock_blueprint_2023}
Langrock,~V.; Krzywda,~J.~A.; Focke,~N.; Seidler,~I.; Schreiber,~L.~R.; Cywinski,~L. Blueprint of a {Scalable} {Spin} {Qubit} {Shuttle} {Device} for {Coherent} {Mid}-{Range} {Qubit} {Transfer} in {Disordered} \$\{{\textbackslash}text\{{Si}/{SiGe}/{SiO}\}\}\_\{2\}\$. \emph{PRX Quantum} \textbf{2023}, \emph{4}, 020305\relax
\mciteBstWouldAddEndPuncttrue
\mciteSetBstMidEndSepPunct{\mcitedefaultmidpunct}
{\mcitedefaultendpunct}{\mcitedefaultseppunct}\relax
\EndOfBibitem
\bibitem[Zwerver \latin{et~al.}(2023)Zwerver, Amitonov, De~Snoo, Mądzik, Rimbach-Russ, Sammak, Scappucci, and Vandersypen]{zwerver_shuttling_2023}
Zwerver,~A.; Amitonov,~S.; De~Snoo,~S.; Mądzik,~M.; Rimbach-Russ,~M.; Sammak,~A.; Scappucci,~G.; Vandersypen,~L. Shuttling an {Electron} {Spin} through a {Silicon} {Quantum} {Dot} {Array}. \emph{PRX Quantum} \textbf{2023}, \emph{4}, 030303\relax
\mciteBstWouldAddEndPuncttrue
\mciteSetBstMidEndSepPunct{\mcitedefaultmidpunct}
{\mcitedefaultendpunct}{\mcitedefaultseppunct}\relax
\EndOfBibitem
\bibitem[Losert \latin{et~al.}(2024)Losert, Oberl\"ander, Teske, Volmer, Schreiber, Bluhm, Coppersmith, and Friesen]{losert2024strategies}
Losert,~M.~P.; Oberl\"ander,~M.; Teske,~J.~D.; Volmer,~M.; Schreiber,~L.~R.; Bluhm,~H.; Coppersmith,~S.; Friesen,~M. Strategies for Enhancing Spin-Shuttling Fidelities in $\mathrm{Si}$/$\mathrm{Si}$$\mathrm{Ge}$ Quantum Wells with Random-Alloy Disorder. \emph{PRX Quantum} \textbf{2024}, \emph{5}, 040322\relax
\mciteBstWouldAddEndPuncttrue
\mciteSetBstMidEndSepPunct{\mcitedefaultmidpunct}
{\mcitedefaultendpunct}{\mcitedefaultseppunct}\relax
\EndOfBibitem
\bibitem[Paquelet~Wuetz \latin{et~al.}(2022)Paquelet~Wuetz, Losert, Koelling, Stehouwer, Zwerver, Philips, Mądzik, Xue, Zheng, Lodari, Amitonov, Samkharadze, Sammak, Vandersypen, Rahman, Coppersmith, Moutanabbir, Friesen, and Scappucci]{paquelet2022atomic}
Paquelet~Wuetz,~B. \latin{et~al.}  Atomic fluctuations lifting the energy degeneracy in Si/SiGe quantum dots. \emph{Nature Communications} \textbf{2022}, \emph{13}\relax
\mciteBstWouldAddEndPuncttrue
\mciteSetBstMidEndSepPunct{\mcitedefaultmidpunct}
{\mcitedefaultendpunct}{\mcitedefaultseppunct}\relax
\EndOfBibitem
\bibitem[Losert \latin{et~al.}(2023)Losert, Eriksson, Joynt, Rahman, Scappucci, Coppersmith, and Friesen]{losert2023practical}
Losert,~M.~P.; Eriksson,~M.~A.; Joynt,~R.; Rahman,~R.; Scappucci,~G.; Coppersmith,~S.~N.; Friesen,~M. Practical strategies for enhancing the valley splitting in Si/SiGe quantum wells. \emph{Phys. Rev. B} \textbf{2023}, \emph{108}, 125405\relax
\mciteBstWouldAddEndPuncttrue
\mciteSetBstMidEndSepPunct{\mcitedefaultmidpunct}
{\mcitedefaultendpunct}{\mcitedefaultseppunct}\relax
\EndOfBibitem
\bibitem[Degli~Esposti \latin{et~al.}(2024)Degli~Esposti, Stehouwer, Gül, Samkharadze, Déprez, Meyer, Meijer, Tryputen, Karwal, Botifoll, Arbiol, Amitonov, Vandersypen, Sammak, Veldhorst, and Scappucci]{degliesposti2024low}
Degli~Esposti,~D. \latin{et~al.}  Low disorder and high valley splitting in silicon. \emph{npj Quantum Information} \textbf{2024}, \emph{10}\relax
\mciteBstWouldAddEndPuncttrue
\mciteSetBstMidEndSepPunct{\mcitedefaultmidpunct}
{\mcitedefaultendpunct}{\mcitedefaultseppunct}\relax
\EndOfBibitem
\bibitem[Lodari \latin{et~al.}(2019)Lodari, Tosato, Sabbagh, Schubert, Capellini, Sammak, Veldhorst, and Scappucci]{lodari_light_2019}
Lodari,~M.; Tosato,~A.; Sabbagh,~D.; Schubert,~M.~A.; Capellini,~G.; Sammak,~A.; Veldhorst,~M.; Scappucci,~G. Light effective hole mass in undoped {Ge}/{SiGe} quantum wells. \emph{Physical Review B} \textbf{2019}, \emph{100}, 041304\relax
\mciteBstWouldAddEndPuncttrue
\mciteSetBstMidEndSepPunct{\mcitedefaultmidpunct}
{\mcitedefaultendpunct}{\mcitedefaultseppunct}\relax
\EndOfBibitem
\bibitem[Degli~Esposti \latin{et~al.}(2022)Degli~Esposti, Paquelet~Wuetz, Fezzi, Lodari, Sammak, and Scappucci]{degli_esposti_wafer-scale_2022}
Degli~Esposti,~D.; Paquelet~Wuetz,~B.; Fezzi,~V.; Lodari,~M.; Sammak,~A.; Scappucci,~G. Wafer-scale low-disorder {2DEG} in {28Si}/{SiGe} without an epitaxial {Si} cap. \emph{Applied Physics Letters} \textbf{2022}, \emph{120}, 184003\relax
\mciteBstWouldAddEndPuncttrue
\mciteSetBstMidEndSepPunct{\mcitedefaultmidpunct}
{\mcitedefaultendpunct}{\mcitedefaultseppunct}\relax
\EndOfBibitem
\bibitem[Paquelet~Wuetz \latin{et~al.}(2020)Paquelet~Wuetz, Bavdaz, Yeoh, Schouten, van~der Does, Tiggelman, Sabbagh, Sammak, Almudever, Sebastiano, Clarke, Veldhorst, and Scappucci]{paquelet2020multiplexed}
Paquelet~Wuetz,~B.; Bavdaz,~P.~L.; Yeoh,~L.~A.; Schouten,~R.; van~der Does,~H.; Tiggelman,~M.; Sabbagh,~D.; Sammak,~A.; Almudever,~C.~G.; Sebastiano,~F.; Clarke,~J.~S.; Veldhorst,~M.; Scappucci,~G. Multiplexed quantum transport using commercial off-the-shelf CMOS at sub-kelvin temperatures. \emph{npj Quantum Information} \textbf{2020}, \emph{6}\relax
\mciteBstWouldAddEndPuncttrue
\mciteSetBstMidEndSepPunct{\mcitedefaultmidpunct}
{\mcitedefaultendpunct}{\mcitedefaultseppunct}\relax
\EndOfBibitem
\bibitem[Monroe \latin{et~al.}(1993)Monroe, Xie, Fitzgerald, Silverman, and Watson]{monroe_comparison_1993}
Monroe,~D.; Xie,~Y.~H.; Fitzgerald,~E.~A.; Silverman,~P.~J.; Watson,~G.~P. Comparison of mobility-limiting mechanisms in high-mobility $\mathrm{Si}_{1-x}\mathrm{Ge}_{x}$ heterostructures. \emph{Journal of Vacuum Science \& Technology B: Microelectronics and Nanometer Structures Processing, Measurement, and Phenomena} \textbf{1993}, \emph{11}, 1731--1737\relax
\mciteBstWouldAddEndPuncttrue
\mciteSetBstMidEndSepPunct{\mcitedefaultmidpunct}
{\mcitedefaultendpunct}{\mcitedefaultseppunct}\relax
\EndOfBibitem
\bibitem[Venkataraman \latin{et~al.}(1993)Venkataraman, Liu, and Sturm]{venkataraman_alloy_1993}
Venkataraman,~V.; Liu,~C.~W.; Sturm,~J.~C. Alloy scattering limited transport of two-dimensional carriers in strained $\mathrm{Si}_{1-x}\mathrm{Ge}_{x}$ quantum wells. \emph{Applied Physics Letters} \textbf{1993}, \emph{63}, 2795--2797\relax
\mciteBstWouldAddEndPuncttrue
\mciteSetBstMidEndSepPunct{\mcitedefaultmidpunct}
{\mcitedefaultendpunct}{\mcitedefaultseppunct}\relax
\EndOfBibitem
\bibitem[Huang and Das~Sarma(2024)Huang, and Das~Sarma]{huang_understanding_2024}
Huang,~Y.; Das~Sarma,~S. Understanding disorder in silicon quantum computing platforms: {Scattering} mechanisms in {Si}/{SiGe} quantum wells. \emph{Physical Review B} \textbf{2024}, \emph{109}, 125405\relax
\mciteBstWouldAddEndPuncttrue
\mciteSetBstMidEndSepPunct{\mcitedefaultmidpunct}
{\mcitedefaultendpunct}{\mcitedefaultseppunct}\relax
\EndOfBibitem
\bibitem[Wuetz \latin{et~al.}(2020)Wuetz, Losert, Tosato, Lodari, Bavdaz, Stehouwer, Amin, Clarke, Coppersmith, Sammak, Veldhorst, Friesen, and Scappucci]{paquelet2020effect}
Wuetz,~B.~P.; Losert,~M.~P.; Tosato,~A.; Lodari,~M.; Bavdaz,~P.~L.; Stehouwer,~L.; Amin,~P.; Clarke,~J.~S.; Coppersmith,~S.~N.; Sammak,~A.; Veldhorst,~M.; Friesen,~M.; Scappucci,~G. Effect of Quantum Hall Edge Strips on Valley Splitting in Silicon Quantum Wells. \emph{Phys. Rev. Lett.} \textbf{2020}, \emph{125}, 186801\relax
\mciteBstWouldAddEndPuncttrue
\mciteSetBstMidEndSepPunct{\mcitedefaultmidpunct}
{\mcitedefaultendpunct}{\mcitedefaultseppunct}\relax
\EndOfBibitem
\bibitem[Goswami \latin{et~al.}(2007)Goswami, Slinker, Friesen, McGuire, Truitt, Tahan, Klein, Chu, Mooney, van~der Weide, Joynt, Coppersmith, and Eriksson]{goswami_controllable_2007}
Goswami,~S.; Slinker,~K.~A.; Friesen,~M.; McGuire,~L.~M.; Truitt,~J.~L.; Tahan,~C.; Klein,~L.~J.; Chu,~J.~O.; Mooney,~P.~M.; van~der Weide,~D.~W.; Joynt,~R.; Coppersmith,~S.~N.; Eriksson,~M.~A. Controllable valley splitting in silicon quantum devices. \emph{Nature Physics} \textbf{2007}, \emph{3}, 41--45\relax
\mciteBstWouldAddEndPuncttrue
\mciteSetBstMidEndSepPunct{\mcitedefaultmidpunct}
{\mcitedefaultendpunct}{\mcitedefaultseppunct}\relax
\EndOfBibitem
\end{mcitethebibliography}

\begin{tocentry}
\includegraphics{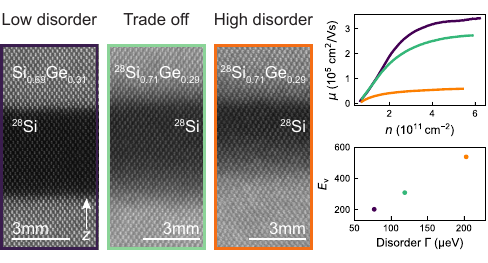}
A  silicon quantum well with diffused interface offers a beneficial trade-off in two-dimensional gases between enhanced valley splitting and electron mobility .
\end{tocentry}

\end{document}


\newpage

\renewcommand{\thefigure}{S\arabic{figure}}

\section{Heterostructure growth}
We grow the Si/SiGe heterostructures in an ASM Epsilon 2000 reduced pressure chemical vapour deposition reactor (RP-CVD). Each heterostructure is grown on a 100 mm n-type Si(001) wafers and begins with a step graded $\mathrm{Si_\mathrm{1-x}}\mathrm{Ge}_\mathrm{x}$ virtual substrate (VS) grown at 750 \textdegree C where the Ge content is graded to a final percentage of 31\% in four steps ($x=0.07,\,0.14,\,0.21,\,0.31$). For heterostructure A, we subsequently lower the growth temperature to 625 \textdegree C to grow a $\mathrm{Si_\mathrm{0.69}}\mathrm{Ge}_\mathrm{0.31}$ strain-relaxed buffer (SRB) using dichlorosilane and germane as precursor gases. We then grow the isotopically purified $^{28}$Si quantum well (800 ppm residual $^{29}$Si isotopes \cite{xue2021cmos, sabbagh2019quantum, xue2022quantum}) using silane at a temperature of 750 \textdegree C, followed by a 30 nm thick $\mathrm{Si_\mathrm{0.69}}\mathrm{Ge}_\mathrm{0.31}$ spacer layer using the same growth conditions as the SRB. Finally, we grow a thin $^{28}$Si capping layer using silane at a temperature of 750 \textdegree C. 

For heterostructures B1-B3 we use use the same step graded VS as in A but subsequently continue the growth of the $^{28}\mathrm{Si_\mathrm{0.71}}\mathrm{Ge}_\mathrm{0.29}$ SRB, the $^{28}$Si quantum well, the 30 nm $^{28}\mathrm{Si_\mathrm{0.71}}\mathrm{Ge}_\mathrm{0.29}$ spacer, and $^{28}$Si capping layer at a growth temperature of 750 \textdegree C using silane and germane as precursor gases.   

For the fabrication of the Hall bars we use the identical procedure as described in the Methods section of Ref.\,\cite{degliesposti2024low}\,.

\section{(S)TEM}
For structural characterization with (S)TEM, we prepared lamella cross-sections of the quantum well heterostructures by using a Focused Ion Beam (Helios
5UX). Atomically resolved HAADF-STEM data was acquired in a double-corrected Thermo Fisher Spectra 300 microscope operated at 300 kV.

\section{Estimation of Ge concentration profiles}
The Ge concentration profiles presented in Fig.\,1(e) are estimated by combining HAADF-STEM data and secondary ion mass spectroscopy (SIMS) data. The atomic resolution HAADF-STEM images presented in Fig.\,1 provide a high spatial accuracy which we make use of by extracting an intensity profile from them. SIMS provides information about the silicon-germanium composition in the barrier layers of the heterostructures. We use the SiGe composition data (see Supplementary Fig.\,S1) of the SiGe barrier layers to rescale the HAADF-STEM intensity profiles to the Ge concentration profiles. 

\section{H-FET measurements}
Characterisation of the Hall bars is done in a Leiden cryogenics refrigerator with a base temperature of 70 mK. We apply a 100 $\upmu$V source-drain bias and measure the current $I_\mathrm{sd}$, longitudinal voltage $V_\mathrm{xx}$ and transverse voltage $V_\mathrm{xy}$ as a function of gate voltage $V_\mathrm{g}$ and perpendicular magnetic field $B$ using four-probe low-frequency lock-in techniques. We calculate longitudinal $\rho_\mathrm{xx}$ and transverse $\rho_\mathrm{xy}$ resistivity from which we extract the Hall density $n$ at low magnetic fields using $\rho_\mathrm{xy}=B/en$, where $e$ is the electron charge. Mobility is found using $\mu=1/ne\rho_\mathrm{xx}$. At $B=0$ we calculate the conductivity $\sigma_\mathrm{xx}=1/\rho_\mathrm{xx}$ and extract the percolation density $n_\mathrm{p}$ using fitting formula $\sigma_\mathrm{xx}\propto(n-n_\mathrm{p})^{1.31}$. We measure a total of 9, 10, 10, and 8 Hall bars for heterostructures A, B1, B2, and B3 respectively.

\section{Valley splitting simulations}
In Refs.~\cite{paquelet2022atomic, losert2023practical}, it was demonstrated that the expected valley splitting of a disorder-dominated quantum dot is given by
\begin{equation} \label{eq:avg_Ev}
    \bar E_\mathrm{v} = \frac{a_0^2 \Delta E_c}{8 a_\text{dot} \Delta G} \sqrt{\sum_l \psi_\text{env}^4 (z_l) G_l (1 - G_l)}
\end{equation} 
where $a_0 = 0.543$~nm is the Si lattice constant, $\Delta E_c$ is the conduction band offset, $a_\text{dot} = \sqrt{\hbar^2 / m_t E_\text{orb}}$ is the dot radius, $\Delta G \approx 0.3$ is the Ge concentration offset between the quantum well and SiGe barriers, $\psi_\text{env} (z)$ is an envelope  function, $G_l$ is the Ge concentration at atomic layer $l$, and the sum is taken over all atomic layers in the heterostructure.
To compute the average quantum dot valley splittings $E_\text{v}^\text{QD}$ reported in Fig.~4, we use Eq.~\ref{eq:avg_Ev}.
In these calculations, we determine $\psi_\text{env}$ by diagonalizing a virtual crystal Hamiltonian
\begin{equation}
    H_\text{vc} = -\frac{\hbar^2}{2 m_l} \partial_z^2 + U_\text{qw} + U_{z},
\end{equation}
where $m_l = 0.916 m_e$ is the longitudinal effective mass in Si.
The quantum well potential is given by
\begin{equation}
    U_\text{qw}(z) = \Delta E_c \frac{G(z) - G_\mathrm{min}}{\Delta G},
\end{equation}
where $G_\text{min}$ is the minimum Ge concentration in the quantum well, and $G(z)$ are shown in Fig.~1. 
The conduction band offset $\Delta E_c$ is computed following Ref.~\cite{paquelet2022atomic}\,.
The electrostatic potential is given by $U_z = -e E_z z$ for constant vertical electric field $E_z = 1$~mV/nm.

Examining Eq.~\ref{eq:avg_Ev}, we can define the dimensionless quantity
\begin{equation} \label{eq:eta_1}
    \eta_1 = \Delta z \sqrt{ \sum_l \psi_\text{env}^4(z_l) G_l (1 - G_l) },
\end{equation}
where $\Delta z = a_0 / 4$ is the spacing between layers in the heterostructure, such that $\bar E_\mathrm{v} \propto \eta_1$.
This quantity captures the overlap of the quantum dot wavefunction into regions with non-zero Ge concentration and is thus a metric for the impact of alloy disorder on the system.
To determine $\psi_\text{env}$ for a Hall bar, we self-consistently solve the Schr\"odinger and Poisson equations. 
The Schr\"odinger-Poisson virtual crystal Hamiltonian is
\begin{equation}
    H_\text{vc}^\text{sp} = -\frac{\hbar^2}{2 m_l} \partial_z^2 + U_\text{qw}  - e \phi(z)
\end{equation}
where the electrostatic potential $\phi(z)$ is determined from Poisson's equation,
\begin{equation}
    \partial_z^2 \phi(z) = -\frac{\rho(z)}{\epsilon} = -\frac{|\psi_\text{env}(z)|^2}{\epsilon}
\end{equation}
where $\epsilon = \epsilon_\text{Si} \epsilon_0$, and $\epsilon_\text{Si} = 11.4$ is the dielectric constant of Si.
To bound these simulations, we choose a total electron density of $1.5 \times 10^{11}$ $\mathrm{cm}^{-2}$, and we enforce $\partial_z \phi(z) = 0$ far below the quantum well. 

\newpage
\begin{figure}
    \centering
	\includegraphics[width=88mm]{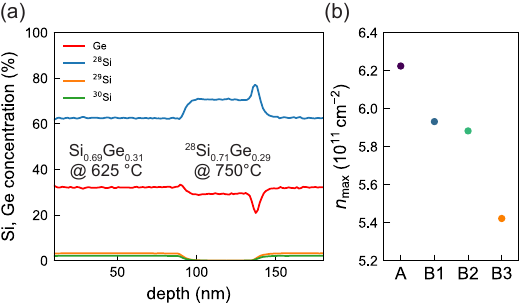}
	\caption{\textbf{Chemical composition and saturation density.} (a) Secondary ion mass spectroscopy (SIMS) of a test structure for the growth of SiGe with isotopically enriched $^{28}$Si. The test structure consists of a layer of $^{28}\mathrm{Si}_{0.71}\mathrm{Ge}_{0.29}$ grown at 750 \textdegree C sandwiched between two layers of all natural $\mathrm{Si}_{0.69}\mathrm{Ge}_{0.31}$ grown at 625 \textdegree C. We use the same growth conditions at 750 \textdegree C of $^{28}\mathrm{Si}_{0.71}\mathrm{Ge}_{0.29}$ for the SiGe barrier layers in heterostructures B1, B2, and B3 of the main text. Heterostructure A uses the growth conditions of the natural $\mathrm{Si}_{0.69}\mathrm{Ge}_{0.31}$ at 625 \textdegree C. (b) Saturation density $n_\mathrm{max}$ for the different heterostructures. We find that B1--B3 show a lower maximum density compared to A, which we attribute to the reduced Ge composition of the barrier layers in these heterostructures. B3 shows an even lower maximum density which most likely is the result of having Ge throughout the entire quantum well in this heterostructure.}
\label{fig:S1}
\end{figure}

\begin{figure}
    \centering
	\includegraphics[width=88mm]{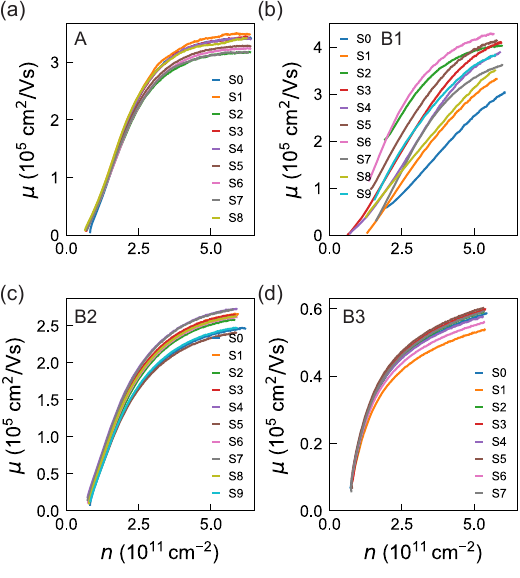}
	\caption{\textbf{Mobility-density characterisation.} (a)-(d) We present the mobility-density results measured across multiple H-FETs (S0-S9) on each heterostructure (A, B1-B3). Heterostructures A, B2, and B3 show uniform mobility-density curves across multiple H-FETs. B1 shows relatively less uniformity, which we attribute to possible strain-relaxation of the quantum well due to a combination of increased growth temperature of the SiGe barrier layers and a comparatively thick $^{28}$Si quantum well of 9.5 nm.}
\label{fig:S2}
\end{figure}

\begin{figure}
    \centering
	\includegraphics[width=88mm]{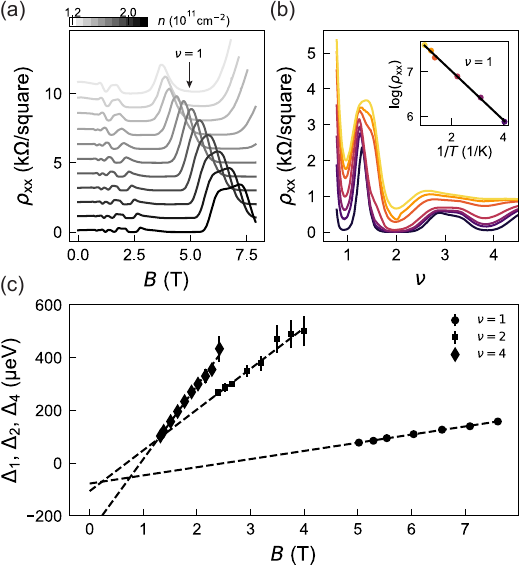}
	\caption{\textbf{Magnetotransport measurements for heterostructure A.} (a) We measure longitudinal resistivity $\rho_\mathrm{xx}$ as a function of magnetic field $B$ for fixed densities $n$ represented by the different gray colours. (b) For each density $n$ (here $1.28\times10^{11}$ cm$^{-2}$) we repeat the magnetotransport measurements at different temperatures $T$ (70-1000 mK) and plot them against integer filling factor $\nu=nh/eB$. At each integer filling factor $\nu$ we extract the corresponding value of $\rho_\mathrm{xx}$ and find a thermally activated dependency given by $\rho_\mathrm{xx}\propto \exp{-\Delta/2k_\mathrm{B}T}$ (see inset), from which we extract the mobility gap $\Delta$. Here we extract $\Delta_\mathrm{v}$ for $\nu=1$ corresponding to the first valley gap. (c) We plot the mobility gaps of the first valley gap $\Delta_\mathrm{v}$ ($\nu=1$), the first Zeeman gap $\Delta_\mathrm{z}$ ($\nu=2$), and the first Landau gap $\Delta_\mathrm{L}$ ($\nu=4$) as function of magnetic field $B$. From a linear fit (dotted line) we extract the Landau level broadening induced disorder $\Gamma$.}
\label{fig:S3}
\end{figure}

\begin{figure}
    \centering
	\includegraphics[width=88mm]{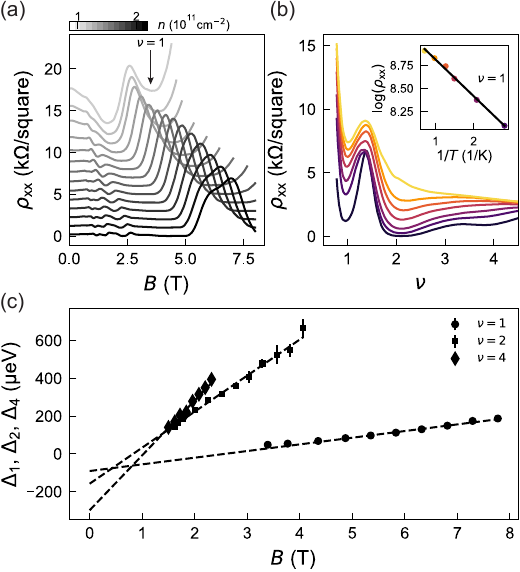}
	\caption{\textbf{Magnetotransport measurements for heterostructure B1.} (a) We measure longitudinal resistivity $\rho_\mathrm{xx}$ as a function of magnetic field $B$ for fixed densities $n$ represented by the different gray colours. (b) For each density $n$ (here $1.05\times10^{11}$ cm$^{-2}$) we repeat the magnetotransport measurements at different temperatures $T$ (70-1000 mK) and plot them against integer filling factor $\nu=nh/eB$. At each integer filling factor $\nu$ we extract the corresponding value of $\rho_\mathrm{xx}$ and find a thermally activated dependency given by $\rho_\mathrm{xx}\propto \exp{-\Delta/2k_\mathrm{B}T}$ (see inset), from which we extract the mobility gap $\Delta$. Here we extract $\Delta_\mathrm{v}$ for $\nu=1$ corresponding to the first valley gap. (c) We plot the mobility gaps of the first valley gap $\Delta_\mathrm{v}$ ($\nu=1$), the first Zeeman gap $\Delta_\mathrm{z}$ ($\nu=2$), and the first Landau gap $\Delta_\mathrm{L}$ ($\nu=4$) as function of magnetic field $B$. From a linear fit (dotted line) we extract the Landau level broadening induced disorder $\Gamma$.}
\label{fig:S4}
\end{figure}

\begin{figure}
    \centering
	\includegraphics[width=88mm]{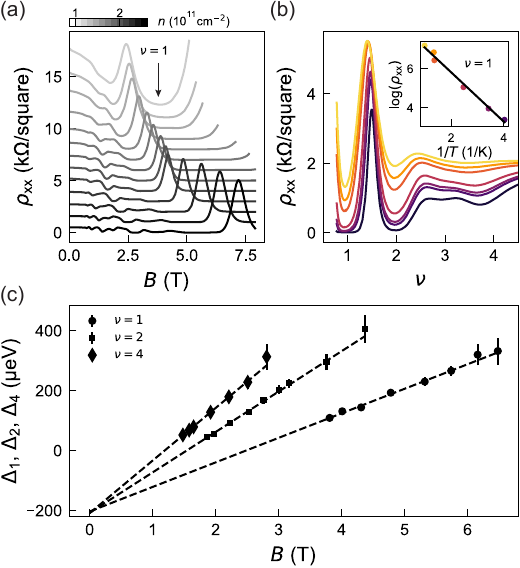}
	\caption{\textbf{Magnetotransport measurements for heterostructure B3.} (a) We measure longitudinal resistivity $\rho_\mathrm{xx}$ as a function of magnetic field $B$ for fixed densities $n$ represented by the different gray colours. (b) For each density $n$ (here $1.21\times10^{11}$ cm$^{-2}$) we repeat the magnetotransport measurements at different temperatures $T$ (70-1000 mK) and plot them against integer filling factor $\nu=nh/eB$. At each integer filling factor $\nu$ we extract the corresponding value of $\rho_\mathrm{xx}$ and find a thermally activated dependency given by $\rho_\mathrm{xx}\propto \exp{-\Delta/2k_\mathrm{B}T}$ (see inset), from which we extract the mobility gap $\Delta$. Here we extract $\Delta_\mathrm{v}$ for $\nu=1$ corresponding to the first valley gap. (c) We plot the mobility gaps of the first valley gap $\Delta_\mathrm{v}$ ($\nu=1$), the first Zeeman gap $\Delta_\mathrm{z}$ ($\nu=2$), and the first Landau gap $\Delta_\mathrm{L}$ ($\nu=4$) as function of magnetic field $B$. From a linear fit (dotted line) we extract the Landau level broadening induced disorder $\Gamma$.}
\label{fig:S5}
\end{figure}

\clearpage

\bibliography{achemso-demo}